# Design A Family of 2D Nb-Based Multilayer Kagome Semimetals with High Fermi Velocity and Low Thermal Conductivity

En-Qi Bao,[1,#] Xing-Yu Wang,[1,2,#] Su-Yang Shen,[1] Jun-Hui Yuan,[1,*] Wen-Yu Fang,[3] and Jiafu Wang,[1,2]

[1]School of Physics and Mechanics, Wuhan University of Technology, Wuhan 430070, China

[2]School of Materials and Microelectronics, Wuhan University of Technology, Wuhan 430070, China

[3]School of Mathematics and Statistics, Hubei University of Science and Technology, Xianning 437100, China

[*]**Corresponding Author**

E-mail: yuanjh90@163.com (J.-H. Yuan)

E.-Q. Bao and X.-Y. Wang contribute equally to this work.

**ABSTRACT**

Although two-dimensional (2D) multilayer kagome materials have opened up new windows of opportunity for exploring novel physical properties, their development has been constrained by the scarcity of available material systems. In light of this, in this study, relying on our previously proposed innovative "1+3" design strategy for multilayer kagome materials, we have successfully designed nine stable 2D niobium-based multilayer kagome monolayers with tunable compositions: $Nb_6Cl_2S_3Br_6$, $Nb_6Cl_2S_4Br_6$, $Nb_6Cl_2Se_3Br_6$, $Nb_6Cl_2Se_4Br_6$, $Nb_6Cl_2S_1Se_3Br_6$, $Nb_6Cl_2S_3Se_1Br_6$, $Nb_6S_4Cl_8$, $Nb_6Se_4Br_8$, and $Nb_6Br_2S_3Se_1Cl_6$. These nine new materials all belong to the category of Dirac semimetals, with their Dirac cone structures primarily arising from the $d_{z^2}$ orbitals based on Nb-based kagome lattice. Hybrid functional calculations reveal that these materials boast Fermi velocities as high as $2.36 \sim 3.04 \times 10^5$ m/s. Moreover, these materials generally exhibit characteristics of relatively low phonon group velocities and shorted phonon lifetimes. Under room temperature conditions, they possess comparatively low lattice thermal conductivities, with values ranging from $1.704 \sim 8.149$ $Wm^{-1}K^{-1}$. Our research not only robustly confirms the feasibility of the "1+3" multilayer kagome lattices design strategy in the realm of kagome material development but also sets an exemplary benchmark for the study of Nb-based multilayer kagome materials.

**Keywords:** Kagome lattice; Two-dimensional materials; Semimetal; Dirac cone; First-principles calculations

# 1. Introduction

Two-dimensional (2D) materials, as a cutting-edge frontier at the intersection of condensed matter physics and materials science, exhibit transformative application potential across various fields such as electronics, photonics, and magnetism, owing to their unique layered architecture and pronounced quantum confinement effects[1–5]. Since the successful exfoliation of graphene in 2004[6] sparked a research boom in 2D materials, systems such as transition metal dichalcogenides and hexagonal boron nitride have been successively discovered[7–10]. However, these materials still fall short in exploring strongly correlated physical phenomena and topological quantum states. Against this backdrop, 2D kagome lattice materials have gradually emerged as an ideal platform for studying strongly correlated physics, topological quantum states, and unconventional superconducting mechanisms, thanks to their distinctive geometric configuration and electronic band characteristics[11–16].

The kagome lattice is intricately woven from vertex-sharing hexagram arrays, forming a periodic network that incorporates triangular and hexagonal sublattices. When confined to two dimensions, this structure exhibits extraordinary electronic band properties, including the coexistence of flat bands, Dirac points, and van Hove singularities[17–21]: flat bands arise from kinetic energy quenching due to electron localization, Dirac points are ingeniously generated by phase interference between sublattices, and van Hove singularities are closely related to the divergence of the density of states at band saddle points. These unique properties provide favorable conditions for realizing novel quantum states such as high-Chern-number topological insulators, quantum spin liquids, and fractional quantum Hall effects[22–25].

Recently, Nb-based kagome materials have garnered significant attention due to their inherent van der Waals layered structure characteristics. A multitude of Nb-based 2D kagome materials, such as $Nb_3X_8$(X=Cl, Br, I)[26–29], and $Nb_3XY_7$ (X = S, Se, Te; Y = Cl, Br, I)[30,31], can be readily obtained through simple mechanical exfoliation methods. These materials have sparked widespread interest in the fields of condensed matter physics and materials science, becoming a research hotspot in the area. For instance, researchers have discovered that in Nb-based "breathing" kagome lattices, triangular sublattices can alternately expand and contract, forming a dynamic breathing mode that endows the material with unprecedented structural degrees of freedom[30]. Furthermore, researchers have unveiled the strong coupling and synergistic effects among "structural breathing, ferroelectricity, and valley degrees of freedom," and successfully achieved electric-field-controlled conversion between multiple valley states using a breathing ferroelectric mechanism, laying a solid theoretical foundation for realizing fully electrically controlled valley electronic devices in breathing kagome lattices[32]. At the device application level, based on the breathing kagome characteristics of $Nb_3Cl_8$, researchers have successfully developed a reconfigurable bipolar field-effect transistor[33]. This transistor enables dynamic switching among various conductive states such as PN, NP, PP, and NN through dual-gate modulation and has demonstrated its application in convolutional image processing, showcasing immense potential in the fields of neuromorphic computing and efficient parallel processing.

The success of 2D Nb-based kagome materials has greatly stimulated research enthusiasm in this field. Compared to materials containing only a single Nb kagome layer, such as $Nb_3X_8$ (X=Cl, Br, I), and $Nb_3XY_7$ (X = S, Se, Te; Y = Cl, Br, I), whether introducing more kagome layers will bring new opportunities has become a question worthy of in-depth exploration. In this work, based on the previously proposed "1+3" strategy for designing multilayer kagome materials[34–36] and combined with stability screening, we successfully obtained nine Nb-based 2D materials with nested multilayer kagome layers. These materials exhibit excellent Dirac cone characteristics at the Fermi level and possess extremely high Fermi velocities, offering new opportunities for the development of high-speed, low-power electronic devices.

## 2. Design Strategy

In this work, the design of Nb-based kagome materials relies on the "1+3" multilayer kagome materials design strategy proposed in our previous research. Based on this strategy, two classic configurations can be obtained, namely the "6+12" configuration and the "6+11" configuration, with their structures shown in **Figures 1a** and **1b**, respectively. Among them, the "6+11" configuration is formed by introducing ordered vacancies into the "6+12" configuration. During the research process, we imposed symmetry constraints on both configurations to make them conform to the $P\bar{3}m1$ space group. Through symmetry analysis, it is found that the non-metal sites can be divided into four non-equivalent sites, namely R, A, X, and D. Based on this, the "6+12" configuration can be expressed as $Nb_6R_2A_3X_1D_6$ according to the stoichiometric ratio, while the "6+11" configuration can be expressed as $Nb_6R_2A_3D_6$. When different non-metal elements are filled at the R, A, X, and D sites, a variety of niobium-based multi-kagome layer materials with different element combinations can be obtained. In this study, only chalcogen elements S/Se and halogen elements Cl/Br are considered as candidates for non-metal elements.

From the perspective of lattice points in the crystal structure, both the "6+12" and "6+11" configurations consist of seven atomic layers, and their skeletons are niobium-based kagome lattices, as shown in **Figure 1c**. However, unlike perfect kagome lattices, the Nb-based kagome lattices in these two configurations exhibit the structural characteristics of breathing kagome lattices, that is, the two triangular lattices have different sizes. Nevertheless, unlike the two triangular lattices in breathing kagome lattices, which can transform into each other through a "breathing" motion, due to symmetry constraints, the two Nb-based kagome lattices in this work cannot undergo such a "breathing" transformation. Therefore, it is more appropriate to define them as "distorted" kagome lattices.

Between the Nb-based skeletons, there are triangular lattices located in the central layer of the crystal, as shown in **Figure 1d**. According to symmetry, this central layer contains X at the center of the hexagon and a kagome lattice formed by A. When ordered X vacancies are introduced, a perfect kagome lattice formed by A can be obtained, as shown in **Figure 1e**. The configuration obtained at this time is the "6+11" configuration.

Thus, the "6+11" configuration forms a spatial structure composed of two layers of "distorted" Nb-based kagome lattices and a central layer of A, which has three kagome layers, as shown in **Figure 1f**. In addition, according to the bonding mode of Nb-D, the outer D atoms also form "distorted" kagome lattices similar to the Nb-based kagome lattice, as shown in **Figure 1g**. Therefore, from the perspective of sublattice distribution, the "6+12" and "6+11" configurations have a total of five kagome layers, as shown in **Figure 1h**.

It should be noted that whether these kagome lattices can exhibit corresponding kagome band characteristics in their electronic structures still requires further analysis. The subsequent calculations in this work will be carried out using the VASP software package[37,38], and the specific calculation settings are detailed in **Note 1** of the **Supplementary Material**.

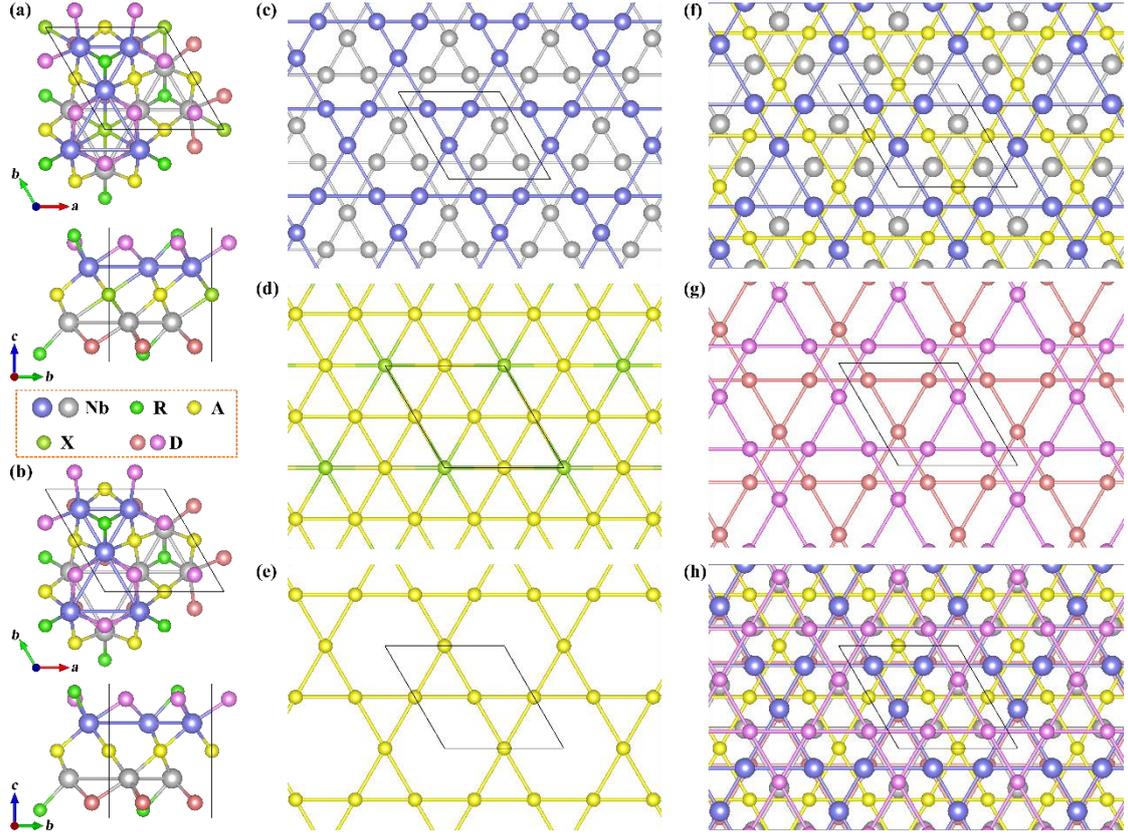

**Figure 1.** Schematic illustrations of the design concept for Nb-based multilayer kagome monolayers. Top and side view of the Nb-based (a) "6+12" and (b) "6+11" crystal model. Top view of (c) the Nb-based bilayer kagome lattice, (d) sublattice of triangle lattice in the center layer, (e) kagome lattice composed of "A" site atoms, (f) trilayer kagome lattice of "6+11" model, (g) the kagome lattice composed of "D" site atoms, (h) the five kagome lattices of "6+11" model.

## 3. Results and discussions

Based on the design strategy described in Section 2, we performed a screening of niobium-based multilayer kagome materials. First, regarding kinetic stability, as shown in **Figs. 1a**, **1b**, and **S1**, nine materials exhibit no imaginary frequency in their phonon spectra, namely: $Nb_6Cl_2S_3Br_6$, $Nb_6Cl_2S_4Br_6$, $Nb_6Cl_2Se_3Br_6$, $Nb_6Cl_2Se_4Br_6$, $Nb_6Cl_2S_1Se_3Br_6$, $Nb_6Cl_2S_3Se_1Br_6$, $Nb_6Cl_2S_4Cl_6$ (abbreviated as $Nb_6S_4Cl_8$),

Nb$_6$Br$_2$Se$_4$Br$_6$ (abbreviated as Nb$_6$Se$_4$Br$_8$), and Nb$_6$Br$_2$S$_3$Se$_1$Cl$_6$. Subsequent *ab initio* molecular dynamics (AIMD) simulations further confirm that these nine novel niobium-based multilayer kagome materials possess excellent thermal stability at room temperature, as illustrated in **Fig. S2**. Finally, mechanical stability is verified via independent elastic constants and the Born–Huang stability criteria[39], as listed in **Table S1**. The validation of kinetic, thermal, and mechanical stabilities demonstrates that these nine new materials are theoretically capable as free-standing 2D materials, thereby indicating considerable research potential.

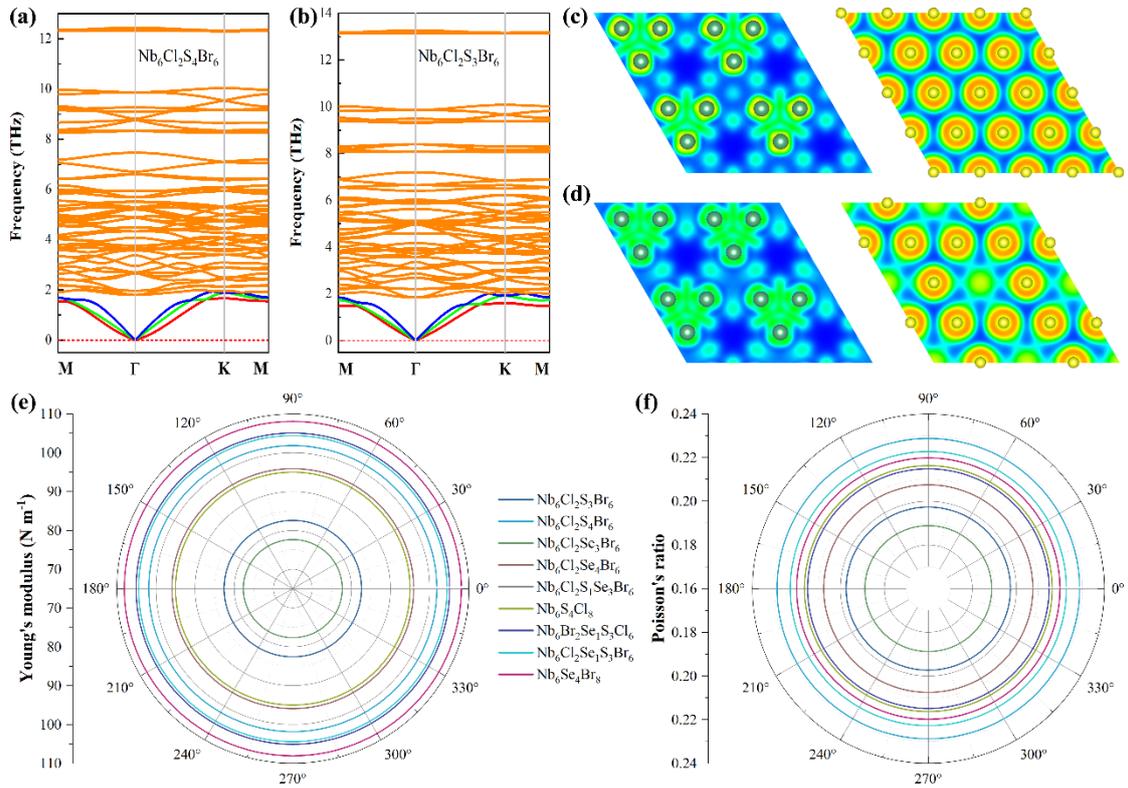

**Figure 2.** Phonon dispersion of (a) Nb$_6$Cl$_2$S$_4$Br$_6$ and (b) Nb$_6$Cl$_2$S$_3$Br$_6$. Electron localization function results for the corresponding Nb atomic layers and S atomic layers in (c) Nb$_6$Cl$_2$S$_4$Br$_6$ and (d) Nb$_6$Cl$_2$S$_3$Br$_6$. The calculated angle-dependent in-plane (e) Young's modulus and (f) Poisson's ratio of Nb-based multilayer kagome monolayer.

The lattice parameters and bond lengths of all nine Nb-based multilayer kagome monolayers exhibit obvious compositional tunability, as summarized in **Table 1**. The lattice constants $a/b$ range from 6.815 Å ($Nb_6S_4Cl_8$) to 7.143 Å ($Nb_6Se_4Br_8$), and the atomic thickness $h$ varies between 6.738 Å ($Nb_6Cl_2S_4Br_6$) and 7.316 Å ($Nb_6Se_4Br_8$), which is attributed to the difference in atomic radii of chalcogen (S/Se) and halogen (Cl/Br) elements: the larger atomic radius of Se/Br leads to lattice expansion compared with S/Cl substitution.

**Table 1.** Calculated lattice constant $a/b$ (Å), atomic thickness $h$ (Å), and bond length $l$ (Å) of Nb-based kagome monolayers.

| Materials | $a/b$ | $h$ | $l_{Nb-Nb}$ | $l_{Nb-R}$ | $l_{Nb-A}$ | $l_{Nb-X}$ | $l_{Nb-D}$ | $l_{A-A}$ | $l_{D-D}$ |
|---|---|---|---|---|---|---|---|---|---|
| $Nb_6Cl_2S_3Br_6$ | 6.922 | 6.779 | 2.870/4.052 | 2.474 | 2.416 | \ | 2.672 | 3.461 | 3.471 |
| $Nb_6Cl_2S_4Br_6$ | 6.980 | 6.738 | 2.869/4.111 | 2.454 | 2.428 | 2.840 | 2.694 | 3.490 | 3.502 |
| $Nb_6Cl_2Se_3Br_6$ | 7.026 | 7.081 | 2.912/4.113 | 2.481 | 2.544 | \ | 2.685 | 3.513 | 3.515 |
| $Nb_6Cl_2Se_4Br_6$ | 7.124 | 7.028 | 2.919/4.205 | 2.476 | 2.549 | 2.964 | 2.714 | 3.562 | 3.561 |
| $Nb_6Cl_2S_1Se_3Br_6$ | 7.080 | 6.993 | 2.931/4.149 | 2.474 | 2.539 | 2.933 | 2.703 | 3.540 | 3.545 |
| $Nb_6Cl_2S_3Se_1Br_6$ | 7.030 | 6.776 | 2.857/4.173 | 2.456 | 2.440 | 2.875 | 2.707 | 3.515 | 3.523 |
| $Nb_6S_4Cl_8$ | 6.815 | 6.819 | 2.860/3.955 | 2.460 | 2.423 | 2.781 | 2.550 | 3.408 | 3.361 |
| $Nb_6Se_4Br_8$ | 7.143 | 7.316 | 2.929/4.214 | 2.598 | 2.543 | 2.960 | 2.706 | 3.571 | 3.562 |
| $Nb_6Br_2S_3Se_1Cl_6$ | 6.908 | 7.191 | 2.868/4.041 | 2.601 | 2.439 | 2.824 | 2.563 | 3.454 | 3.390 |

A typical structural feature of all monolayers is the dual-value distribution of Nb-Nb bond lengths (2.857~2.931 Å/3.955~4.214 Å), which is the hallmark of a distorted kagome lattice (trimerization) and distinguishes it from the perfect kagome lattice with uniform Nb-Nb bond lengths. This distortion is induced by the symmetry constraints of the *P*-3*m*1 space group and the selective filling of non-metal elements at different lattice sites, and it cannot undergo the "breathing transformation" of traditional breathing kagome lattices, forming a new type of Nb-based distorted kagome lattice structure. The Nb-nonmetal bond lengths (Nb-R, Nb-A, Nb-X, Nb-D) are concentrated in 2.416~2.964 Å, and the interatomic distances of nonmetal sublattices ($l_{A-A}$, $l_{D-D}$) are in 3.361~3.571 Å, forming a stable covalent bonding framework between Nb skeletons and nonmetal atoms, which provides the structural basis for the stability of the multilayer kagome structure.

The electron localization function (ELF)[40] was used to analyze the electron distribution and bonding nature of the representative $Nb_6Cl_2S_4Br_6$ and $Nb_6Cl_2S_3Br_6$, as shown in **Figs. 2c** and **2d**. The results show that the ELF of Nb-based distorted kagome lattices exhibits typical trimerization characteristics. The sub-triangular lattices formed by short Nb–Nb bonds are filled with nearly free electrons (ELF ≈ 0.5), whereas distinct electron vacuum regions appear in the sub-triangular lattices formed by long Nb–Nb bonds (ELF ≈ 0). Meanwhile, a small amount of electrons accumulate along the perpendicular bisectors of the Nb–Nb bonds. In contrast, the most pronounced difference in ELF is observed in the central S layer of the seven-atom stacking structure. In pristine $Nb_6Cl_2S_4Br_6$ without S vacancies, electrons are highly localized around S atoms, and the regions between S–S bonds are electron-depleted. Upon introducing S vacancies, the electrons in the central S layer of $Nb_6Cl_2S_3Br_6$ undergo redistribution: on the one hand, electrons in the remaining kagome lattice of S atoms show a tendency to accumulate; on the other hand, a small number of nearly free electrons are still trapped in the vacancy sites under the influence of the lattice potential. These free electrons can still form bonds with Nb, allowing the system to approximately retain the electronic structure properties of $Nb_6Cl_2S_4Br_6$. This conclusion is further confirmed by subsequent electronic structure calculations.

We evaluated the mechanical stability of nine materials based on independent elastic constants (results are shown in **Table S2**). All materials met the Born-Huang criteria[39], indicating reliable structural mechanical stability. **Figs. 2e** and **2f** display the angle-dependent in-plane Young's modulus and Poisson's ratio calculated from independent elastic constants, revealing isotropic in-plane elastic stiffness in the materials. The Young's modulus values ranged from 77.63 to 105.08 N m$^{-1}$, falling into the medium-to-high modulus category. $Nb_6Br_2S_3Se_1Cl_6$ had the highest Young's modulus, while $Nb_6Cl_2Se_3Br_6$ had the lowest, with a difference of 27.45 N m$^{-1}$, highlighting the significant impact of composition regulation on stiffness. The Poisson's ratio values ranged from 0.189 to 0.229, exhibiting low Poisson's ratio characteristics with minimal fluctuations (a maximum difference of 0.04). The transverse deformation response was less influenced by composition regulation compared to Young's modulus, with $Nb_6Cl_2S_4Br_6$ having the highest and $Nb_6Cl_2Se_3Br_6$ having the lowest Poisson's ratio. Further analysis revealed a clear pattern in material performance under elemental composition regulation: The S/Se ratio was the primary regulatory factor. Due to its smaller atomic radius and higher electronegativity, sulfur forms shorter and higher-energy bonds with niobium, enhancing bonding strength and in-plane stiffness, and causing a slight increase in Poisson's ratio, demonstrating an S-substitution hardening effect. Thus, an increase in sulfur content significantly improved both Young's modulus and Poisson's ratio. Chlorine/bromine halogen substitution was a secondary regulatory factor, with bromine introduction increasing modulus while having minimal impact on Poisson's ratio. The synergistic ratio of S/Se and Cl/Br allowed for continuous adjustment of Young's modulus while maintaining a low and stable Poisson's ratio. Additionally, the symmetry of the kagome lattice resulted in significant in-plane mechanical isotropy in the materials.

The electronic band structure of the Nb-based kagome monolayers was calculated by both GGA-PBE and hybrid functional HSE06 methods. As shown in **Fig. 3a** and **S3**, the results show that all nine materials are intrinsic Dirac semimetals. Taking $Nb_6Cl_2S_3Br_6$ as a typical example, the band structure exhibits a perfect Dirac cone structure exactly at the Fermi level, and the band crossing point is not accompanied by band gap opening in both GGA-PBE and HSE06 calculations. The HSE06 method, which better describes the exchange-correlation potential, further confirms the robustness of the Dirac cone, eliminating the possibility of band gap opening caused by calculation method errors. Moreover, the calculated results from HSE06 demonstrate that the energy bands near the Dirac cone exhibit a steeper linear dispersion.

As shown in **Fig. 3b**, **3c** and **S4-S10**, projected band structure analysis reveals the orbital origin of the Dirac cone: the electronic states at the Fermi level are primarily derived from the $d_{z^2}$ orbitals of the Nb-based kagome lattice, while the contribution of nonmetal (S/Se/Cl/Br) orbitals is very limited. This indicates that the Nb atomic skeleton is the core of the electronic properties of the materials near the Fermi level, and the chalcogen/halogen elements only play a regulatory role in the band dispersion rather than changing the intrinsic Dirac semimetal characteristics. This structural-electronic property correlation ensures the universality of Dirac semimetal characteristics in this material family, and the compositional substitution will not destroy the Dirac cone structure at the Fermi level.

As shown in **Fig. 4a**, the 3D band structure of $Nb_6Cl_2S_3Br_6$ intuitively shows the conical band dispersion at the Fermi level, and the linear dispersion relationship near the Dirac point is the key to the high Fermi velocity of the materials. The 3D energy bands near the Dirac cone for the other eight materials are shown in **Fig. S11**. The linear band dispersion means that the electron effective mass is close to the bare electron mass, which is beneficial for the rapid transport of charge carriers in the material.

Fermi velocity is a key parameter for evaluating the charge transport performance of semimetallic materials, and we calculated the Fermi velocities of the nine monolayers along the Γ-K and K-M high-symmetry directions based on GGA-PBE and HSE06 methods (**Fig. 4b** and **Table S2**). The results show that the Fermi velocities of all materials are in the range of $2.36×10^5$ to $3.04×10^5$ m/s, which is comparable to that of graphene (~$8×10^5$ m/s)[41], demonstrating excellent charge transport performance.

The Fermi velocity of the materials exhibits obvious compositional tunability: S/Cl substitution leads to a higher Fermi velocity than Se/Br substitution. For example, $Nb_6S_4Cl_8$ with pure S/Cl nonmetal sublattice has the highest Fermi velocity (~$3.04×10^5$ m/s), while Nb$_6$Se$_4$Br$_8$ with pure Se/Br substitution has a relatively low Fermi velocity (~2.36×10$^5$ m/s). This is because the S/Cl substitution results in a more compact lattice and a steeper linear band dispersion near the Dirac point, which reduces the electron effective mass and thus increases the Fermi velocity. In addition, the Fermi velocity along the Γ-K direction is slightly higher than that along the K-M direction for all materials, reflecting the weak in-plane anisotropy of charge transport, which is consistent with the hexagonal lattice symmetry of the $P\bar{3}m1$ space group.

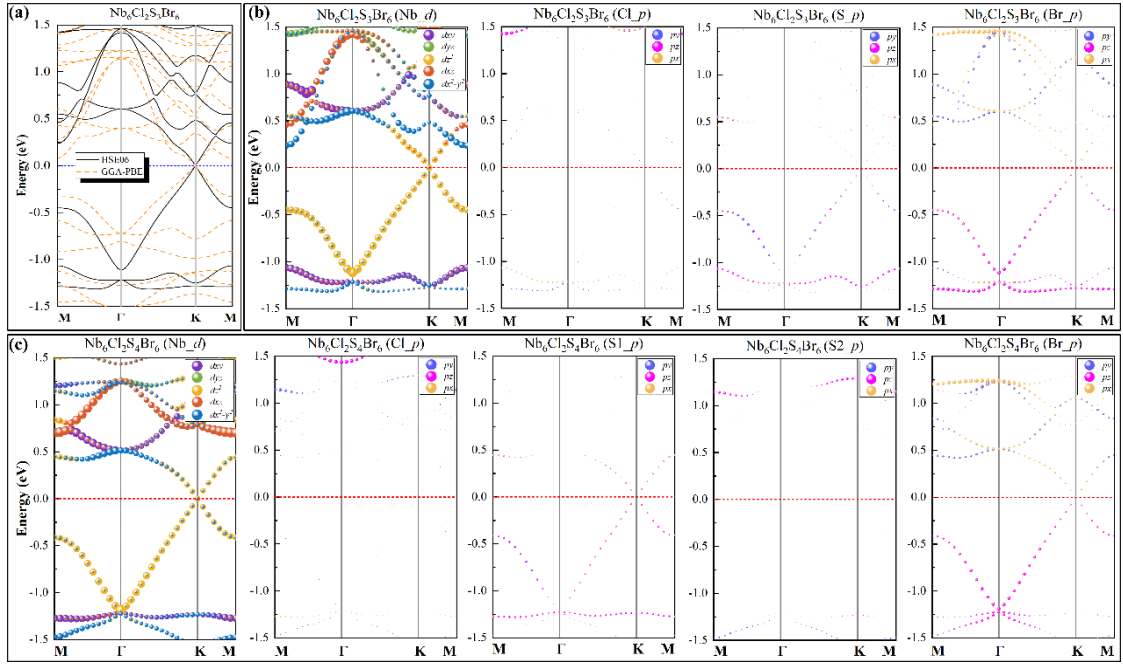

**Figure 3.** (a) Band structure of Nb$_6$Cl$_2$S$_3$Br$_6$ based on GGA-PBE and HSE06 calculations. Projected band structures of (b) Nb$_6$Cl$_2$S$_3$Br$_6$ and (c) Nb$_6$Cl$_2$S$_4$Br$_6$.

The high and tunable Fermi velocity of this material family originates from the synergistic effect of the distorted Nb kagome lattice and the linear band dispersion of $d_{z^2}$ orbitals: the delocalized Nb-$d_{z^2}$ electrons form a continuous charge transport channel, and the linear band dispersion near the Dirac point ensures the low effective mass of charge carriers. This unique electronic structure makes the Nb-based multilayer kagome monolayers promising candidates for high-speed, low-power nanoelectronic devices.

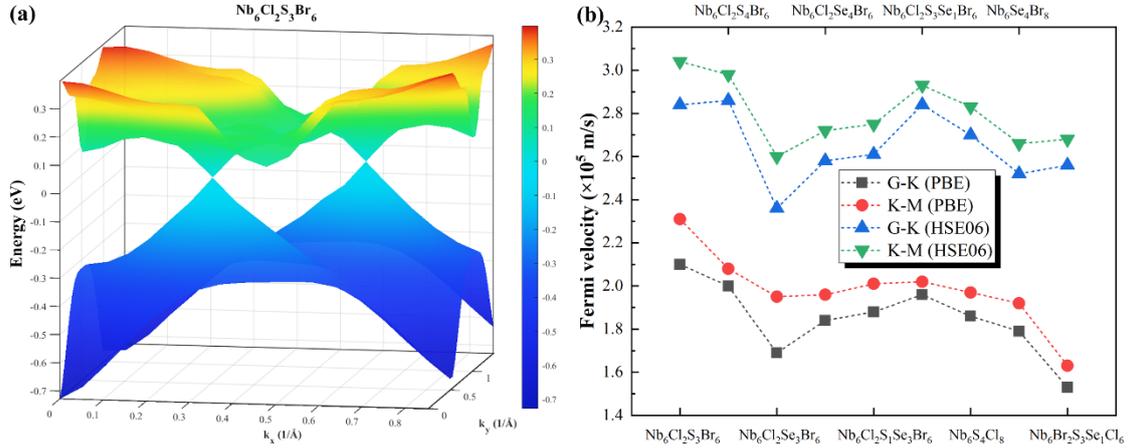

**Figure 4.** (a) 3D band structures of $Nb_6Cl_2S_3Br_6$ monolayers based on GGA-PBE calculations. (b) Calculated Fermi velocity of Nb-based kagome monolayers based on GGA-PBE and HSE06 calculations.

Thermal transport properties are closely related to phonon dynamics, and we first calculated the phonon group velocity and phonon lifetime of the representative $Nb_6Cl_2S_3Br_6$ at room temperature, as shown in **Figs. 5a** and **5b**. The results of the phonon group velocity and phonon lifetime for the other eight Nb-based kagome materials are presented in **Figs. S12** and **S13**. The phonon group velocities of the Nb-based kagome monolayers are concentrated in the range of 0 to 4.6 km/s, and the average group velocity is significantly lower than that of traditional 2D materials (e.g., graphene ~22 km/s[42], phosphorene ~8 km/s[43]). The low phonon group velocity is mainly attributed to the distorted kagome lattice structure: the asymmetric Nb-Nb bond lengths and the weak coupling between Nb skeletons and nonmetal layers lead to the softening of acoustic phonon branches and the reduction of phonon propagation speed.

The phonon lifetime of the materials at room temperature is mainly in the range of 0 to 1000 ps, and the optical phonon lifetime is significantly shorter than the acoustic phonon lifetime (~250 ps). The shortening of phonon lifetime originates from the strong phonon scattering probability within the material: the structure composed of seven atomic layers stacked together, along with the non-uniform non-metallic sublattice, enhances the probability of phonon-phonon scattering as well as acoustic-optical phonon scattering, thereby resulting in a relatively short phonon lifetime. Notably, the coexistence of low phonon group velocity and short phonon lifetime is a unique feature of this material family, which jointly determines its low lattice thermal conductivity characteristics.

Lattice thermal conductivity is the core parameter of thermal transport properties, and we calculated the temperature-dependent lattice thermal conductivity of all nine Nb-based kagome monolayers from 100 K to 600 K (**Fig. 5c**). The results show that the lattice thermal conductivity of these nine materials increases with the increase of temperature, and the growth rate gradually slows down, which conforms to the classic thermal transport law of 2D materials: the lattice thermal conductivity is dominated by phonon-phonon scattering at low temperature, and the scattering probability increases with temperature, leading to the slow growth of thermal conductivity. At room temperature, the lattice thermal conductivity of the nine monolayers ranges from 1.704 ($Nb_6S_4Cl_8$) to 8.149 ($Nb_6Cl_2Se_3Br_6$) $Wm^{-1}K^{-1}$, which is much lower than that of traditional 2D materials (e.g., graphene ~5300 $Wm^{-1}K^{-1}$,[44] $MoS_2$ ~54 $Wm^{-1}K^{-1}$[45]), demonstrating excellent low thermal conductivity characteristics.

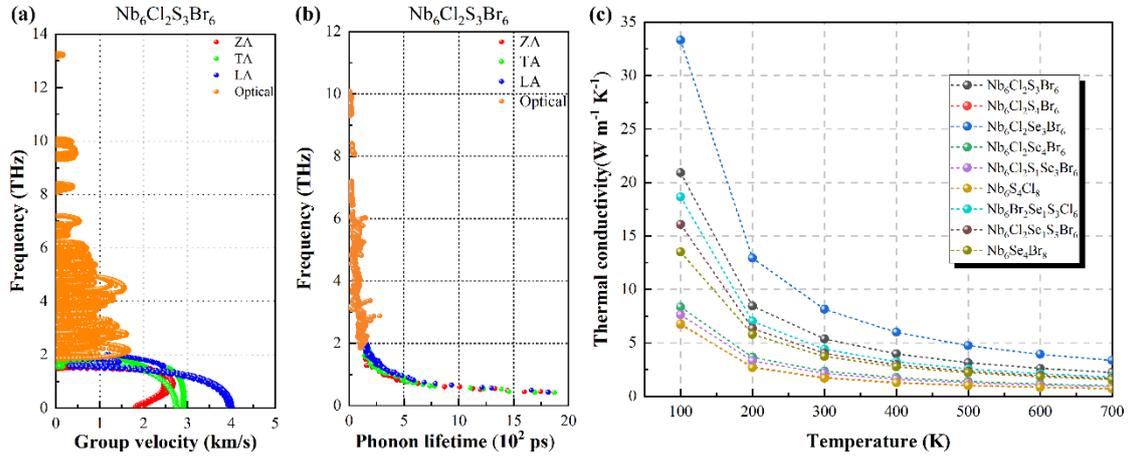

**Figure 5.** (a) Phonon group velocity and (b) phonon lifetime of $Nb_6Cl_2S_3Br_6$ monolayers at room temperature. (c) Temperature-dependent lattice thermal conductivity of Nb-based kagome monolayers.

The low lattice thermal conductivity of the Nb-based multilayer kagome monolayers is the result of the combined action of three factors: the distorted kagome lattice leads to low phonon group velocity, the seven-atomic-layer stacked structure reduces phonon propagation efficiency, and the compositional inhomogeneity of nonmetal sublattices enhances phonon scattering. This unique thermal transport characteristic, combined with the high Fermi velocity, makes the Nb-based multilayer kagome materials promising candidates for high-performance thermoelectric materials, as the low lattice thermal conductivity can effectively reduce the thermal loss and improve the thermoelectric figure of merit.

The systematic characterization of the nine Nb-based multilayer kagome monolayers reveals a clear structure-property correlation: the distorted kagome lattice structure of the Nb atomic skeleton determines the intrinsic Dirac semimetal characteristics and high Fermi velocity, while the selective filling of chalcogen/halogen elements at nonmetal lattice sites realizes the tunability of structural, mechanical, electronic and thermal transport properties. The "1+3" multilayer kagome material design strategy proposed in our previous work is fully verified in this material family: by dividing the nonmetal lattice sites into four non-equivalent sites (R, A, X, D) and filling different chalcogen/halogen elements, a series of stable 2D Nb-based kagome materials with tunable properties can be designed. Notably, the design strategy breaks through the limitation of single kagome layer in traditional Nb-based kagome materials (e.g., $Nb_3X_8$), and constructs a multilayer kagome lattice structure with five nested kagome layers, which brings new physical properties such as high Fermi velocity and low lattice thermal conductivity. This work not only expands the family of 2D Nb-based kagome materials but also provides a universal design method for the development of multilayer kagome materials: the combination of the "1+3" strategy and compositional substitution can realize the precise regulation of material properties, which is of great significance for the exploration of novel 2D topological quantum materials and their applications in nanoelectronics and thermoelectrics.

## 4. Conclusion

Based on the "1+3" design strategy and first-principles calculations, nine stable 2D Nb-based multilayer kagome monolayers with compositional tunability were designed. Their kinetic, thermal and mechanical stability were fully verified, confirming theoretical feasibility as free-standing 2D materials. All materials are intrinsic Dirac semimetals with Dirac cones at the Fermi level (derived from Nb-$d_{z^2}$ orbitals) and high Fermi velocities (2.36~3.04×$10^5$ m/s) with compositional tunability. They also exhibit ultra-low lattice thermal conductivity (1.704~8.149 $Wm^{-1}K^{-1}$ at room temperature) due to the distorted lattice, seven-atomic-layer structure and nonmetal sublattice inhomogeneity. This work validates the "1+3" strategy, expands 2D Nb-based kagome materials, and provides promising candidates for nanoelectronic and thermoelectric devices, with the design method extendable to other transition metal-based multilayer kagome materials.

**Acknowledgements**

This work was supported by the Fundamental Research Funds for the Central Universities (WUT: 2024IVA052), the National Undergraduate Innovation and Entrepreneurship Training Program (Grant No. 202510497080). We would like to express our gratitude for the support from the High-performance Computing Platform of Wuhan University of Technology.